\documentclass[sigconf]{acmart}
\pdfoutput=1
\usepackage{booktabs} 
\usepackage{verbatim}
\usepackage[nolist]{acronym}

\usepackage[utf8x]{inputenc}
\usepackage{listings}

\usepackage{algorithm2e}
\usepackage{multirow}
\usepackage{subcaption}
\usepackage{tcolorbox}
\usepackage{etoolbox}

\setcopyright{acmcopyright}

\acmDOI{10.475/123_4}

\acmISBN{123-4567-24-567/08/06}

\acmConference[ASP-DAC 2019]{ACM Asia South Pacific Design Automation Conference}{Jan. 2019}{Tokyo, Japan}
\acmYear{2019}
\copyrightyear{2019}

\acmArticle{4}
\acmPrice{15.00}

\editor{Jennifer B. Sartor}
\editor{Theo D'Hondt}
\editor{Wolfgang De Meuter}

\newcommand{\ceil}[1]{\left\lceil #1 \right\rceil}

\begin{document}
\title{Insights into the Mind of a Trojan Designer}
\subtitle{The Challenge to Integrate a Trojan into the Bitstream}

\author{Maik Ender}
\authornote{All three authors contributed equally to the paper}
\orcid{0000-0002-0685-2541}
\affiliation{%
  \institution{Ruhr-Universität Bochum \\Horst Görtz Institute for IT Security}
  \city{Bochum}
  \country{Germany}
  \postcode{44801}
}
\email{maik.ender@rub.de}

\author{Pawel Swier­czyn­ski}
\authornotemark[1]
\affiliation{%
  \institution{Digital Society Institute, ESMT}
  \city{Berlin}
  \country{Germany}
  \postcode{10178}
}
\email{pawel.swierczynski@esmt.org}

\author{Sebastian Wallat}
\authornotemark[1]
\orcid{0000-0002-7429-1002}
\affiliation{%
  \institution{University of Massachusetts}
  \city{Amherst, MA}
  \country{USA}
  \postcode{01003}
}
\email{swallat@umass.edu}

\author{Matt­hi­as Wil­helm}
\affiliation{%
  \institution{Ruhr-Universität Bochum \\Horst Görtz Institute for IT Security}
  \city{Bochum}
  \country{Germany}
  \postcode{44801}
}
\email{matthias.wilhelm@rub.de}

\author{Paul Martin Knopp}
\affiliation{%
  \institution{Ruhr-Universität Bochum \\Horst Görtz Institute for IT Security}
  \city{Bochum}
  \country{Germany}
  \postcode{44801}
}
\email{paul.knopp@rub.de}

\author{Christof Paar}
\affiliation{%
  \institution{Ruhr-Universität Bochum \\Horst Görtz Institute for IT Security}
  \city{Bochum}
  \country{Germany}
  \postcode{44801}
}
\email{Christof.Paar@rub.de}

\renewcommand{\shortauthors}{Maik Ender, Pawel Swier­czyn­ski, Sebastian Wallat, et. al.}

\begin{abstract}
The threat of inserting hardware Trojans during the design, production, or in-field poses a danger for integrated circuits in real-world applications. A particular critical case of hardware Trojans is the malicious manipulation of third-party FPGA configurations. In addition to attack vectors during the design process, FPGAs can be infiltrated in a non-invasive manner after shipment through alterations of the bitstream. First, we present an improved methodology for bitstream file format reversing. Second, we introduce a novel idea for Trojan insertion.
\end{abstract}

%
%

\ccsdesc[500]{Security and privacy~Hardware reverse engineering}
\ccsdesc[500]{Security and privacy~Malicious design modifications}
\ccsdesc[500]{Security and privacy~Embedded systems security}
\maketitle

\begin{acronym}
 \setlength{\itemsep}{0.2em}
 \acro{RTL}{Register Transfer Level}
 \acro{3DES}{Triple DES}
 \acro{AA}{Active Authentication}
 \acro{ADC}{Analog to Digital Converter}
 \acro{ACK}{Acknowledgement}
 \acro{AES}{Advanced Encryption Standard}
 \acro{AID}{Application Identifier}
 \acro{ANF}{Algebraic Normal Form}
 \acro{APDU}{Application Protocol Data Unit}
 \acro{API}{Application Programming Interface}
 \acro{ASCII}{American Standard Code for Information Interchange}
 \acro{ASCA}{Algebraic Side-Channel Analysis}
 \acro{ASK}{Amplitude-Shift Keying}
 \acro{ASIC}{Application Specific Integrated Circuit}
 \acro{ATQA}{Answer To Request A}
 \acro{ATR}{Answer To Reset}
 \acro{ATS}{Answer To Select}
 \acro{BAC}{Basic Access Control}
 \acro{BPSK}{Binary Phase Shift Keying}
 \acro{BSI}{Bundesamt für Sicherheit in der Informationstechnik}
 \acro{CBC}{Cipher Block Chaining}
 \acro{CC}{Common Criteria}
 \acro{CCA}{Canonical Correlation Analysis}
 \acro{CHES}{Cryptographic Hardware and Embedded Systems}
 \acro{CIA}{Combined Implementation Attacks} 
 \acro{CL}{Cascade Level}
 \acro{CMRR}{Common Mode Rejection Ratio}
 \acro{CMTF}{Combined Masking in Tower Fields}
 \acro{CMOS}{Complementary Metal Oxide Semiconductor}
 \acro{COPACOBANA}{Cost-Optimized Parallel Code Breaker and Analyzer}
 \acro{COSY}{Communication Security}
 \acro{CPA}{Correlation Power Analysis}
 \acro{CPU}{Central Processing Unit}
 \acro{CRC}{Cyclic Redundancy Check}
 \acro{CRT}{Chinese Remainder Theorem}
 \acro{CTR}{Counter \acroextra{(mode of operation)}}
 \acro{DAC}{Digital-Analog Converter}
 \acro{DECT}{Digital Enhanced Cordless Telecommunications}
 \acro{DES}{Data Encryption Standard}
 \acro{DEMA}{Differential Electro-Magnetic Analysis}
 \acro{DC}{Direct Current}
 \acro{DCM} {Digital Clock Manager}
 \acro{DFA}{Differential Frequency Analysis}
 \acro{DFT}{Discrete Fourier Transform}
 \acro{DoM}{Difference of Means}
 \acro{DoS}{Denial-of-Service}
 \acro{DIP}{Dual Inline Package}
 \acro{DIMM}{Dual In Line Memory Modules}
 \acro{DOM}{Difference-of-Means}
 \acro{DPA}{Differential Power Analysis}
 \acro{DRAM}{Dynamic Random Access Memory}
 \acro{DSO}{Digital Storage Oscilloscope}
 \acro{DSP}{Digital Signal Processing}
 \acro{DST}{Digital Signature Transponder}
 \acro{DTW}{Dynamic Time Warping}
 \acro{DUT}{Device Under Test}
 \acroplural{DUT}[DUTs]{Devices Under Test}
 \acro{DVB-T}{Digital Video Broadcasting -- Terrestrial}
 \acro{EAC}{Extended Access Control}
 \acro{ECB}{Electronic Code Book}
 \acro{ECC}{Elliptic Curve Cryptography}
 \acro{ECDLP}{Elliptic Curve Discrete Logarithm Problem}
 \acro{ECMA}{European Computer Manufacturers Association}
 \acro{EDE}{Encrypt-Decrypt-Encrypt \acroextra{(mode of operation)}}
 \acro{EEPROM}{Electrically Erasable Programmable Read-Only Memory}
 \acro{EM}{Electro-Magnetic}
 \acro{EMC}{Electro-Magnetic Compatibility}
 \acro{EMSEC}{Embedded Security}
 \acro{EOC}{End Of Communication}
 \acro{EOF}{End Of File}
 \acro{ePass}{Electronic Passport}
 \acro{EPC}{Electronic Product Code}
 \acro{EU}{European Union}
 \acro{FDT}{Frame Delay Time}
 \acro{FF}{Flip Flop}
 \acro{FFT}{Fast Fourier Transform}
 \acro{FI}{Fault Injection}
 \acro{FIB}{Focused Ion Beam}
 \acro{FIFO}{First In First Out \acroextra{(memory)}}
 \acro{FIPS}{Federal Information Processing Standard}
 \acro{FIR}{Finite Impulse Response} 
 \acro{FIT}{Faculty of Information Technology}
 \acro{FPGA}{Field Programmable Gate Array}
 \acro{FSM}{Finite-State Machine}
 \acro{FSK}{Frequency Shift Keying}
 \acro{GIAnT}{Generic Implementation Analysis Toolkit}
 \acro{GND}{Ground}
 \acro{GNFS}{General Number Field Sieve}
 \acro{GUI}{Graphical User Interface}
 \acro{GPIO}{General Purpose I/O}
 \acro{GPL}{GNU General Public License}
 \acro{GPS}{Global Positioning System}
 \acro{GSM}{Global System for Mobile Communications}
 \acro{HAC}{Handbook of Applied Cryptography}
 \acro{HD}{Hamming Distance}
 \acro{HDL}{Hardware Description Language}
 \acro{HF}{High Frequency}
 \acro{HGI}{Horst Görtz Institute for IT Security}
 \acro{HID}{Human Interface Device}
 \acro{HLTA}{HaLT type A}
 \acro{HMAC}{Hash-based Message Authentication Code}
 \acro{HOTP}{HMAC-based One Time Password}
 \acro{HVSP}{High Voltage Serial Programming} 
 \acro{HVG}{High Voltage Generator}
 \acro{HW}{Hamming Weight}
 \acro{ICAO}{International Civil Aviation Organization}
 \acro{IC}{Integrated Circuit}
 \acro{ICC}{Integrated Circuit Card}
 \acro{ID}{Identifier}
 \acro{IDE}{Integrated Development Environment}
 \acro{IFD}{Interface Device}
 \acro{IGBT}{Insulated Gate Bipolar Transistor}
 \acro{ISO}{International Organization for Standardization}
 \acro{ISM}{Industrial, Scientific, and Medical \acroextra{(frequencies)}}
 \acro{IFF}{Identify Friend or Foe}
 \acro{IIR}{Infinite Impulse Response}
 \acro{IP}{Intellectual Property}
 \acro{ISR}{Interrupt Service Routine}
 \acro{ISP}{In-System Programming}
 \acro{IV}{Initialization Vector}
 \acro{JTAG}{Joint Test Action Group}
 \acro{KDF}{Key Derivation Function}
 \acro{LAN}{Local Area Network}
 \acro{LED}{Light-Emitting Diode}
 \acro{LF}{Low Frequency}
 \acro{LFSR}{Linear Feedback Shift Register}
 \acro{LIW}{Listening Window}
 \acro{LSB}{Least Significant Bit}
 \acro{LSByte}{Least Significant Byte}
 \acro{LUT}{Look-Up Table}
 \acro{MAC}{Message Authentication Code}
 \acro{MF}{Medium Frequency}
 \acro{MIA}{Mutual Information Analysis}
 \acro{MITM}{Man-In-The-Middle}
 \acro{MOSFET}{Metal-Oxide Semiconductor Field-Effect Transistor}
 \acro{MRZ}{Machine Readable Zone}
 \acro{MRTD}{Machine Readable Travel Document}
 \acro{MSB}{Most Significant Bit}
 \acro{MSByte}{Most Significant Byte}
 \acro{muC}[$\mathrm{\upmu C}$]{Microcontroller}
 \acro{nPA}{New German ID Card}
 \acro{NACK}{Negative Acknowledgment}
 \acro{NDA}{Non-Disclosure Agreement}
 \acro{NIST}{National Institute of Standards and Technology}
 \acro{NLFSR}{Non-Linear Feedback Shift Register}
 \acro{NLF}{Non-Linear Function}
 \acro{NFC}{Near Field Communication}
 \acro{NRZ}{Non-Return-to-Zero \acroextra{(encoding)}}
 \acro{NOP}{No Operation}
 \acro{NVM}{Non-Volatile Memory}
 \acro{OATH}{Initiative of Open Authentication}
 \acro{OOK}{On-Off-Keying}
 \acro{OP}{Operational Amplifier}
 \acro{OS}{Operating System}
 \acro{OTP}{One-Time Password}
 \acro{PA}{Passive Authentication}
 \acro{PACE}{Password Authenticated Connection Establishment}
 \acro{PC}{Personal Computer}
 \acro{PCA}{Principal Component Analysis}
 \acro{PCB}{Printed Circuit Board}
 \acro{PCD}{Proximity Coupling Device}
 \acro{PEA}{Photonic Emission Analysis}
 \acro{PICC}{Proximity Integrated Circuit Card}
 \acro{PIP}{Programmable Interconnect Point}
 \acro{PIT}{Programmable Identification Transponder}
 \acro{PKI}{Public Key Infrastructure}
 \acro{PLL} {Phase Locked Loop}
 \acro{PMS}{Perfectly Masked Squaring}
 \acro{PMM}{Perfectly Masked Multiplication}
 \acro{PPC}{Pulse Pause Coding}
 \acro{PPS}{Protocol and Parameter Selection}
 \acro{PPSR}{Protocol and Parameter Selection Request}
 \acro{PRN}{Pseudo-Random Number}
 \acro{PRNG}{Pseudo-Random Number Generator}
 \acro{PS}{Passive Serial \acroextra{(mode)}}
 \acro{PSK}{Phase Shift Keying}
 \acro{RADAR}{Radio Detection And Ranging}
 \acro{RAM}{Random Access Memory}
 \acro{RATS}{Request for Answer To Select}
 \acro{REQA}{REQuest type A}
 \acro{RISC}{Reduced Instruction Set Computer}
 \acro{RF}{Radio Frequency}
 \acro{RFID}{Radio Frequency IDentification}
 \acro{RGT}{Request Guard Time}
 \acro{RKE}{Remote Keyless Entry}
 \acro{RNG}{Random Number Generator}
 \acro{ROM}{Read Only Memory}
 \acro{RSA}{Rivest Shamir and Adleman}
 \acro{RTF}{Reader Talks First}
 \acro{RUB}{Ruhr-University Bochum}
 \acro{SAK}{Select AcKnowledge}
 \acro{SAM}{Square-and-Multiply}
 \acro{SCA}{Side-Channel Analysis}
 \acro{SDK}{Software Development Kit}
 \acro{SDR}{Software-Defined Radio}
 \acro{SECT}{Security Transponder}
 \acro{SEM}{Scanning Electron Microscopy}
 \acro{SNR}{Signal to Noise Ratio}
 \acro{SHA}{Secure Hash Algorithm}
 \acro{SHA-1}{Secure Hash Algorithm 1}
 \acro{SMA}{SubMiniature version A \acroextra{(connector)}}
 \acro{SQL}{Structured Query Language}
 \acro{SOF}{Start Of Frame}
 \acro{SOIC}{Small-Outline Integrated Circuit}
 \acro{SPA}{Simple Power Analysis}
 \acro{SPOF}{Single Point of Failure}
 \acro{SRAM}{Static Random Access Memory}
 \acro{TA}{Template Attack}
 \acro{TEM}{Transmission Electron Microscopy}
 \acro{TLU}{Table Look Up}
 \acro{TTF}{Tag Talks First}
 \acro{TMTO}{Time-Memory Tradeoff}
 \acro{TMDTO}{Time-Memory-Data Tradeoff}
 \acro{TMM}{Transformed Multiplicative Masking}
 \acro{TNR}{Trace-to-Noise Ratio}
 \acro{TWI}{Two Wire Interface}
 \acro{SOC}{Start Of Communication}
 \acro{SHF}{Superhigh Frequency}
 \acro{SPI}{Serial Peripheral Interface}
 \acro{UC}[$\mu$C]{Microcontroller}
 \acro{UHF}{Ultra High Frequency}
 \acro{UID}{Unique Identifier} 
 \acro{UMTS}{Universal Mobile Telecommunications System}
 \acro{USB}{Universal Serial Bus}
 \acro{USRP}{Universal Software Radio Peripheral}
 \acro{USRP2}{Universal Software Radio Peripheral (version 2)}
 \acro{UVC} [UV-C] {Ultraviolet-C \acroextra{(light)}}
 \acro{VCP}{Virtual COM Port}
 \acro{VHF}{Very High Frequency}
 \acro{VLF}{Very Low Frequency}
 \acro{VHDL}{VHSIC (Very High Speed Integrated Circuit) Hardware Description Language}
 \acro{WLAN}{Wireless Local Area Network}
 \acro{WUPA}{Wake-UP A}
 \acro{XOR}{Exclusive OR}
\acro{AES}{Advanced Encryption Standard}
\acro{ASIC}{Application Specific Integrated Circuit}
\acro{CPA}{Correlation Power Analysis}
\acro{DPA}{Differential Power Analysis}
\acro{SPA}{Simple Power Analysis}
\acro{DUT}{Device Under Test}
\acro{DFT}{Discrete Fourier Transform}
\acro{DFA}{Differential Frequency Analysis}
\acro{SCA}{Side-Channel Analysis}
\acro{FIR}{Finite Impulse Response}
\acro{FPGA}{Field Programmable Gate Array}
\acro{muC}[$\mathrm{\upmu C}$]{microcontroller}
\acro{SCA}{Side-Channel Analysis}
\acro{USB}{Universal Serial Bus}
\acro{SRAM}{Static Random Access Memory}
\acro{API}{Application Programming Interface}
\acro{EM}{electro-magnetic}
\acro{PCB}{Printed Circuit Board}
\acro{DSO}{Digital Storage Oscilloscope}
\acro{IC}{Integrated Circuit}
\acro{UHF}{Ultra-High Frequency}
\acro{HF}{High Frequency}
\acro{MAC}{Message Authentication Code}
\acro{RNG}{Random Number Generator}
\acro{SNR}{Signal-to-Noise Ratio}
\acro{IV}{Initialization Vector}
\acro{HW}{Hamming Weight}
\acro{HD}{Hamming Distance}
\acro{ECB}{Electronic Code Book}
\acro{CBC}{Cipher Block Chaining}
\acro{CFB}{Cipher Feedback Mode}
\acro{OFB}{Output Feedback Mode}
\acro{CTR}{Counter}
\acro{UART}{Universal Asynchronous Receiver Transmitter}
\acro{DC}{Direct Current}
\acro{IV}{Initialization Vector}
\acro{AES}{Advanced Encryption Standard}
\acro{JTAG}{Joint Test Action Group}
\acro{HDL}{Hardware Description Language}
\acro{PS}{Passive Serial}
\acro{IP}{Intellectual Property}
\acro{AES}{Advanced Encryption Standard}
\acro{3DES}{Triple-DES}
\acro{NVM}{Non-Volatile Memory}
\acro{LFSR}{Linear Feedback Shift Register}
\acro{PC}{Personal Computer}
\acro{LSB}{Least Significant Bit}
\acro{MSB}{Most Significant Bit}
\acro{DSP}{Digital Signal Processing}
\acro{FPGA}{Field Programmable Gate Array}
\acro{HSM}{Hardware Security Module}
\acro{DNF}{Disjunctive Normal Form}
\acro{XDL}{Xilinx Design Language}
\acro{XTS}{XEX Tweakable Block Cipher with Ciphertext Stealing}
\acro{NSA}{National Security Agency}
\acro{BRAM}{Block Random Access Memory}
\acrodefplural{BRAM}[BRAMs]{Block Random Access Memories}

\acro{IOB}{Input Output Block}
\acro{CLB}{Configurable Logik Block}
\acro{DSA}{Digital Signature Algorithm}
\acro{RSA-PSS}{RSA Signature Scheme with Appendix - Probabilistic Signature Scheme}
\acro{CPLD}{Complex Programmable Logic Device}
\acro{SPN}{Substitution-Permutation Network}
\acro{PUF}{Physically Unclonable Function}
\acro{ARM}{Advanced RISC Machine}
\acro{BiFI}{Bitstream Fault Injection}
\acro{OEM}{Original Equipment Manufacturer}

\acro{MUX}{multiplexer}

\end{acronym}

\section{Introduction}\label{aspdac:sec:intro}

The threat of \ac{IP} theft, imposed by hardware reverse engineering, has been historically considered as the main practical security issue.
The move from on-site fabrication to a globally distributed supply-chain and the arising threats of interdiction changed this perspective significantly for all kind of applications.

Since the Snowden's surveillance revelations, malicious hardware manipulations became an increasing concern, including SRAM-based \acp{FPGA}.
Due to the volatile nature of SRAM-based \acp{FPGA}, new attack vectors arise such as bitstream interception and manipulation. A prerequisite for those kinds of attacks is to reverse engineer the bitstream file formats. For this reason and to support highly customisable bitstream generation tools, various research works \cite{JBITS,JPG,PARBITS,4101017,rannaud2008bitstream,6339165,DingWZZ13,7927114,debit,max5,prjxray} aimed at reverse engineering the proprietary bitstream file format of SRAM-based FPGAs, which mainly focused on Xilinx \acp{FPGA}.

However, today it is not possible to fully reverse the entire bitstream format of Xilinx FPGAs which reveals all the details of a specific hardware configuration. Hence, there is no official support for developing open source bitstream generation tools similar to Project IceStorm~\cite{IceStorm}, which reversed the Lattice iCE40 FPGAs. Such a tool improves the flexibility for designers and researchers, i.e., it could extend (security) frameworks like \textit{HAL}~\cite{HAL}, Torc~\cite{DBLP:conf/fpga/SteinerWSCAF11}, or RapidSmith~\cite{DBLP:conf/fpt/LavinPLNH10}.

Knowing the entire bitstream file format, the security of cryptographic hardware configuration can be appropriately analyzed. Thus quick-and-easy malicious bitstream manipulation attacks ~\cite{ROtrojan,aldaya_2015,TCAD,USBK,BIFI}, leading to a potential security breach, can be pentested beforehand and accordingly addressed by a security analyst. Defending of FPGA designs is even more crucial since most FPGA bitstream encryption schemes of older FPGA generations are vulnerable to side-channel attacks~\cite{SCAVII,SCAV4V5,STRII,STRIII,bs_enc_hacked} or do not offer any bitstream encryption/authentication at all. Hence the hardware layout reverse engineering and manipulation of bitstreams are a real threat. Notably, many old systems used in large infrastructures deploy hundreds of (older) FPGA devices. Newer hardware modules cannot simply replace them due to high costs or environmental reasons. Hence, considering the long life span~\cite{AlteraReport15} of (older) deployed SRAM-based FPGAs, it is always worthwhile to explore the practical doability of bitstream reverse engineering, hardware design reverse engineering, and the corresponding potential malicious hardware manipulations, cf. Wallat~\textit{et~al.}~\cite{darkside}. All those methods need to be understood well, as they are crucial for improving the security of critical systems.
Summarizing bitstream reverse engineering can be used for illegitimate and legitimate purposes.

The main issue of bitstream file format reverse engineering and a meaningful hardware configuration manipulation is a seemingly complicated and time-consuming practical task. In general, it is unclear to what extent an attacker can use a non-perfect converted netlist from a bitstream to inject a hardware Trojan into it.

In this work, we provide insights into bitstream reverse engineering techniques and Trojan insertion strategies at the hardware configuration level. Our contribution is as follows.
\begin{enumerate}
    \item We present an improved methodology for bitstream file format reversing targeting its routing encoding. Moreover, it is capable of extracting the bitstream encoding rules for \acp{PIP}, \acp{LUT}, and \acp{FF}, which we exemplarily conducted for a Xilinx Spartan~6 \ac{FPGA}. Additionally,  our framework manipulates bitstreams, e.g., it can replace LUT configurations and set/unset single PIPs.
    \item We introduce a new method for a hardware Trojan insertion into a self-test-protected AES IP core at the hardware configuration level. This gives an idea of how advanced attacker may perform malicious hardware configuration manipulations.
\end{enumerate}

\section{Background}
In this section, we briefly introduce the needed background on SRAM-based \acp{FPGA}, which is necessary to fully comprehend the bitstream reversing and manipulation methods presented within this paper. For an in-depth description of FPGAs, we refer the interested reader to~\cite{DBLP:journals/pieee/TessierPD15}.

\begin{figure}[htb]
    \includegraphics[width=1\columnwidth]{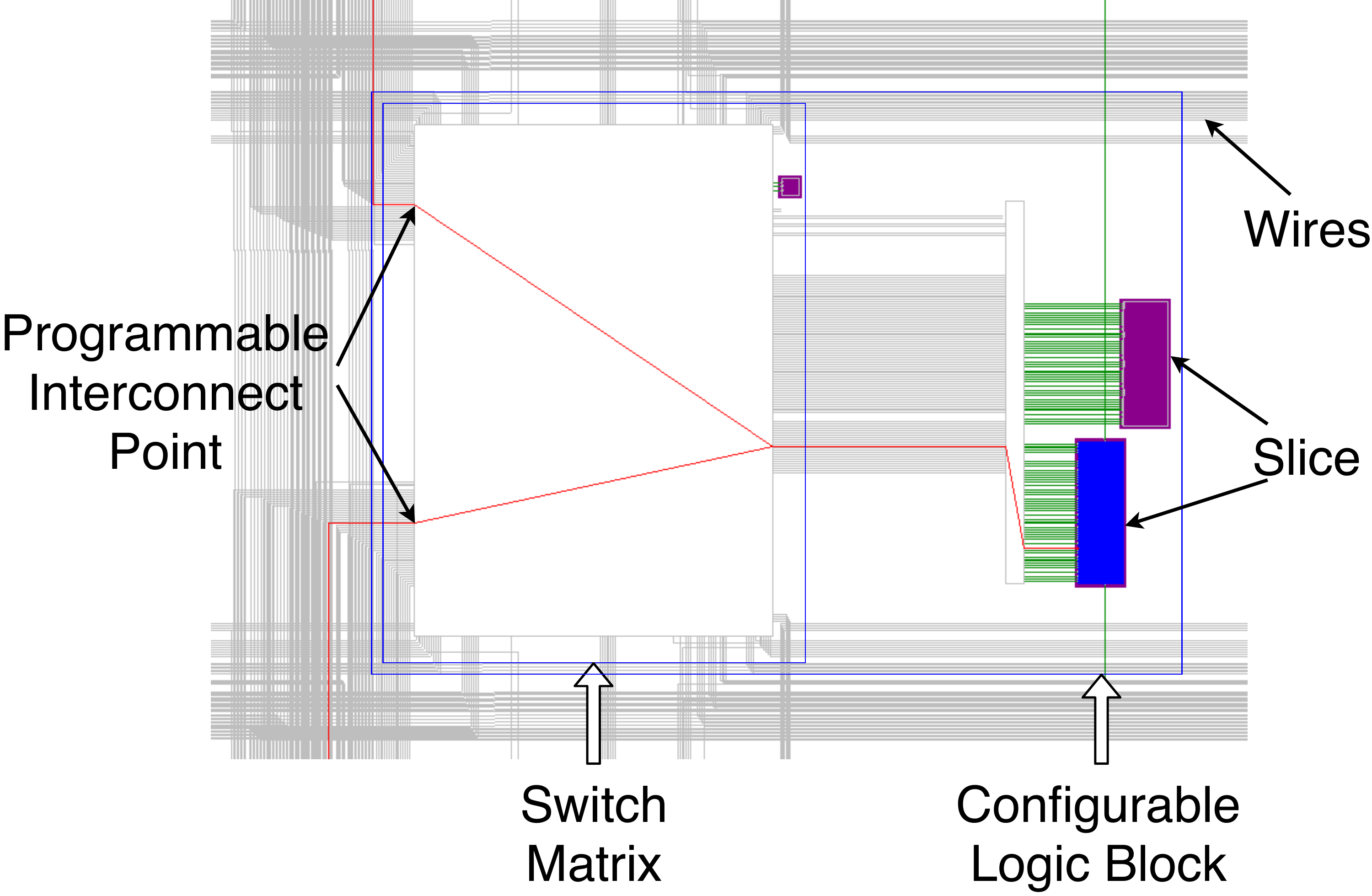}
    \caption{Part of configured FPGA internals showing a switch matrix, a NET, and a SLICE.}
    \label{fig:aspdac:background:FPGA}
\end{figure}

The Spartan-6 has a two-dimensional array structure, where SLICEs encounter the \acp{LUT} and \acp{FF}. Two SLICES and an adjacent switch matrix form one \ac{CLB}, as this is depicted in Figure~\ref{fig:aspdac:background:FPGA}.
The switch matrix realizes the interconnection logic of the FPGA by configuring so-called \acp{PIP}.

As indicated by Figure~\ref{fig:aspdac:background:FPGA}, \acp{PIP} are configurable wires within a switch matrix that connect static wires with other static wires. Hence, they allow building specific bridges between wires. Since the PIPs are reprogrammable, the information must be stored in the proprietary bitstream files. Generally speaking, the PIPs realize the actual routing functionality of an FPGA.

Furthermore, a dedicated signal with a defined source and sink is called a NET. Sources and sinks are for example the LUT's input or FF's output. The PIPs realize the NETs' routing from a dedicated source to its sinks, i.e., an FF's output is routed via several PIPs to a LUT's input. Thus, when knowing all potential PIP configurations, the NETs can be reconstructed.

Finally, we shortly note that the bitstream is an encoded version of the placed and routed XDL netlist file. We refer to this information as hardware configuration.

\section{Bitstream Reverse Engineering}\label{aspdac:sec:bs_rev}
We divide the reverse engineering process into two phases; the (1) bitstream reverse engineering and (2) bitstream conversion phase.

The first phase determines the relation between all bits in the bitstream and the associated hardware configuration of a specific FPGA model. The result of the first step is a database containing the mapping of a single bitstream's bit and its impact on the hardware configuration of a hardware primitive such as \acp{FF}, multiplexers, \acp{PIP}, or \acp{LUT}.

In the second phase, a targeted bitstream is converted back into a human-readable netlist representation by using the previously created database.
The resulting gate-level netlist (or: XDL file) can then be processed by officially supported tools such as the ISE suite and FPGA Editor from Xilinx or by unofficially supported tools like \textit{RapidSmith}~\cite{DBLP:conf/fpt/LavinPLNH10}.
With this step, the before unknown targeted bitstream's hardware configuration is revealed. Thus, it can be for example further processed by a reverse engineer.

The reverse engineering of the information that encodes the routing within a bitstream was already depicted in ~\cite{DingWZZ13}, but we provide an improved methodology for the Xilinx Spartan-6 Series. In particular, we were successful for the \texttt{6slx16csg324} and \texttt{6slx75csg484} FPGAs. Our approach simplifies and speeds up the entire bitstream reverse-engineering process. For reverse-engineering and verification purposes, we used the Xilinx ISE Suite 14.7.

\subsection{Phase~1: Bitstream Reversing}\label{aspdac:sec:bs_rev:phase1}

Our tool for reversing the bitstream creates a database containing the mapping between most configurable FPGA resources and its configuration bits in the bitstream. In the following, we focus on the PIP reversing, but the methodology is also applicable to other FPGA components. Note that we reversed LUTs, flip-flops, and multiplexers as well, but we do not provide futher details. Our approach works as follows.

\begin{enumerate}
    \item Creation of a minimalistic template (XDL netlist) that either configures a single hardware elements, e.g., a PIP.\label{steps:template}
    \item Generation of a reference bitstream which encodes supportive instances regarding the template of Step~\ref{steps:template}.\label{steps:emptyBitstream}
    \item Conducting template alterations followed by bitstream generation and bit toggle observation.\label{steps:reversing}
    \item Database creation.\label{steps:database}
\end{enumerate}

To correlate the configuration of FPGA resources with the bitstream bits, we created specific crafted XDL netlist templates (Step~\ref{steps:template}) that are translated by the vendors tool to a bitstream (Step~\ref{steps:emptyBitstream}). Note that for each FPGA element, an individual template is needed, e.g, PIPs require a different template than LUTs. During Step~\ref{steps:reversing}, the configuration of the templates are slightly modified and another bitstream is generated, again using the vendors tool. The resulting change of both generated bitstream files are recorded into a database (Step~\ref{steps:database}). For bitstream creation, we used the Xilinx tools \textit{xdl} and \textit{bitgen} as~follows.
\begin{itemize}
    \item xdl -xdl2ncd -force empty.xdl // creates an NCD file
    \item bitgen -d empty.ncd // creates the corresponding bitstream
\end{itemize}

By using the \textit{-force} and \textit{-d} option, this enables us to generate bitstreams even though the hardware configuration itself is an illegitimate design.

Note that some targeted hardware elements require the configuration of other hardware elements. Otherwise, the vendor's tools do not manipulate the bitstream files accordingly.
This is an issue that we faced ourselves when we tried to reverse engineer the bitstream encoding for one PIP configuration. It is not sufficient to only configure a PIP within a NET, but one needs to attach the NET to other instances. Listing~\ref{lst:templatePIP} exemplary shows an XDL template that generates a valid bitstream containing information about the PIP configuration.

\vspace{16pt}
\begin{lstlisting}[breaklines=true,basicstyle=\footnotesize\ttfamily,frame=tb,numbers=left, numbersep=5pt, caption=XDL template for one PIP configuration (WW2E2 $\rightarrow$ NL1B2) of the switch matrix INT\_X10Y16, captionpos=b, abovecaptionskip=6pt, label=lst:templatePIP]
design "basic_pip" xc6slx16csg324-3 v3.2;
inst "q" "IOB",placed BIOB_X11Y0 T8;
inst "q_OBUF" "OLOGIC2",placed BIOI_OUTER_X11Y0 OLOGIC_X8Y0;
net "q_OBUF" , 
outpin "q_OBUF" OQ ,
inpin "q" O , 
pip INT_X10Y6 WW2E2 -> NL1B2  ;

\end{lstlisting}
\vspace{14pt}

We first reverse engineer the routing encoding of one switch matrix and later on simplify the process for all other available ones. In our example (cf. Listing~\ref{lst:templatePIP}), we target the switch matrix labeled as INT\_X10Y16. In line~4, a NET is instanciated containing the examined PIP (WW2E2 $\rightarrow$ NL1N2) in line~7. To our surprise, we found out that it is just necessary to specify two \textit{arbitrary chosen} instances (line~2 and~3) that are supposed to be connected with the NET's OUTPIN and INPIN, even if the NET does not route such connectivity. Hence, there is no need to worry about finding and connecting the NET to valid instances, which eases the entire reverse engineering task significantly.

Additionally, thanks to the \textit{-force} and \textit{-d} parameters, the bitstream generation tools nevertheless encode the PIP's information into the resulting bitstream even though our specified instance and NET connectivity are not necessarily correct. If the two arbitrary chosen instances are missing in the XDL template, the PIP's configuration is not encoded into the bitstream file even though the specified NET itself is valid.

Since we instantiate two additional instances which are not associated with the PIP, its configuration is undesirably encoded into every generated bitstream file. This information can be easily removed by letting the vendors tool generate a reference bitstream file (only once) that does only encode both instances (line 2 and 3) without the NET in Listing~\ref{lst:templatePIP}, i.e., Step~\ref{steps:emptyBitstream}.

The difference between any generated bitstream (instances plus NET with individual PIP) and the one-time generated reference bitstream (with instances only) reveals the correct bitstream encoding of an individual PIP configuration. Hence, both bitstreams simply need to be XORed.

Which PIPs are available for a switch-matrix configuration, can be figured out quickly by parsing the information from so-called XDL report files that can be easily generated, cf. ~\cite{6339165}. For every possible PIP the template is changed, e.g., Line~7 is replaced with another examined PIP and the corresponding bitstream is generated.

Further note that in cases where a PIP configuration does not lead to any bit toggle in the bitstream, we mark it as a default configuration.
During our reverse engineering efforts, this only happened for a fraction of all PIPs (under 1\% of all PIPs).

The presented PIP reversing approach of Ding~\textit{et al.}~\cite{DingWZZ13} relies~on~$i)$~creating NETs with multiple PIPs and on $ii)$ conducting further pre- and postprocessing steps. As opposed to that, our approach minimizes the required pre- and postprocessing steps and hence can be carried out faster and in a less complicated manner. It also does require less knowledge about the FPGA internals.

Note that by applying the described approach of~\cite{DingWZZ13}, we can also reduce the time for reversing the bitstream and the file size of the database. This works by exploiting the repetitiveness of hardware elements distributed over the FPGA's grid structure.

Most switch matrices are of the same type, i.e., they contain the same labeled PIPs. Moreover, the byte distances among the configuration bits from different PIPs are always equally distributed within one switch matrix type. Exemplarily, the configuration bits of the first and second PIP could always be separated by $k$ byte positions, which would apply to (most) switch matrices. This was described by  Ding~\textit{et al.}~\cite{DingWZZ13}.

\begin{figure*}[!htb]
	\centering
	\begin{subfigure}[b]{0.49\textwidth}
		\includegraphics[width=\textwidth]{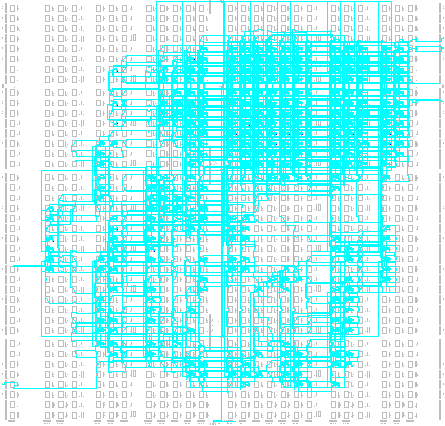}
		\caption{Routing information of the original hardware configuration as it is generated by the vendors tools.}
		\label{fig:gull}
	\end{subfigure}
	\begin{subfigure}[b]{0.49\textwidth}
		\includegraphics[width=\textwidth]{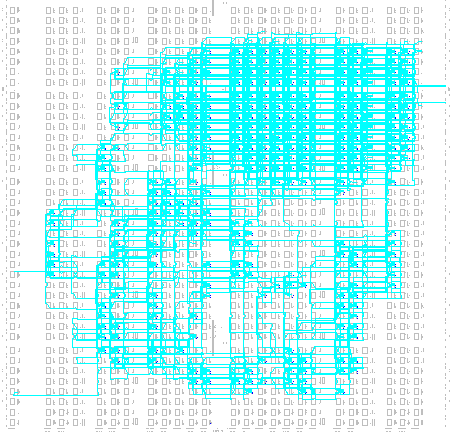}
		\caption{Extracted routing information from the bitstream. The conversion was carried out with our framework.}
		\label{fig:tiger}
	\end{subfigure}
	\caption{Routing comparison between the original and the converted AES IP core.}\label{fig:animals}
	\label{fig:screenshot}
\end{figure*}

We were able to verify this, as we have also reversed various switch matrices with different locations on the FPGA grid for verification purposes. As expected, the bitstream encoding yielded the same distances. Additionally, the routing of all our tested FPGA designs could be later on correctly recovered. A sample design recovery for the routing of an AES IP core is depicted in Figure~\ref{fig:screenshot}. Our recovered netlist only lacks the clock tree information which is work in progress. Otherwise, the recovered routing information is complete.

Once all PIPs of a single switch matrix are reverse engineered (revealing all distances), it is sufficient to only generate one bitstream for every remaining unreversed switch matrix in the FPGA. Each of those bitstreams encodes the configuration of a chosen fixed reference PIP.

To calculate the positions of the remaining unreversed PIPs of a switch matrix, the offset of the fixed reference PIP is added with the corresponding previously derived PIP distance. Each PIP position can be hence simply computed with
\begin{equation}
	PIP\_position~=~reference\_PIP\_position~+~PIP\_distance
\end{equation}
Given that we have to reverse-engineer one switch matrix with $3461$ possible PIPs (distance PIPs) and there are only $2278$ switch matices (reference pips), we just need to generate $5739=3461+2278$ bitstreams to derive a complete list of most available PIPs.

A straight-forward attempt to reverse most PIPs of all switch matrices individually would make the required reverse-engineering time practically infeasible: when considering that a bitstream generation plus its processing (single threaded) takes~$\sim$7 seconds on an average machine (i7-7700HQ, 3.8GHz, 4 cores, 8 threads) and that there are approx.~$\sim$5.5 million possible PIPs on a Spartan-6 (6slx16), the sequential bitstream generation of all~$5.5$ million bitstreams would take approximately~$\sim$381~days.

As opposed to that, it took us only approximately~$21.5$~hours to reverse engineer the entire routing encoding by following the distance approach of Ding. We also implemented a parallelized version with 8 threads and could further reduce the reverse engineering time to~$\sim$2.6~hours.

Once we have generated all bitstreams, where each bitstream encodes the configuration of a chosen fixed reference PIP for one switch-matrix, we derive a database (Step~\ref{steps:database}) that stores all those reference byte positions. Additionally, we store the distance patterns for all PIPs that can later on be used to reconstruct the exact byte and bit positions  (Equation~1) for any queried switch matrix.

\subsection{Phase~2: Bitstream Conversion}\label{aspdac:sec:bs_rev:phase2}
\begin{figure}[!htb]
    \centering
    \includegraphics[width=\columnwidth]{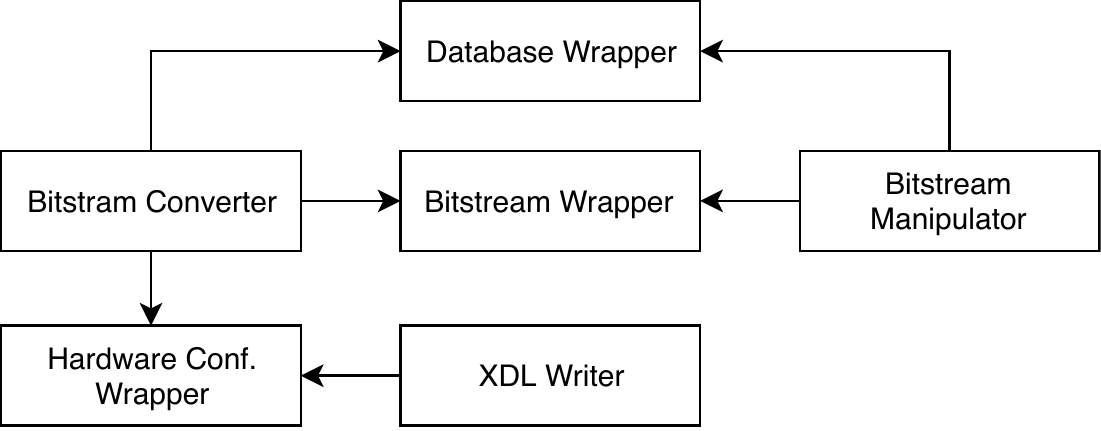}
    \caption{Architecture of our bitstream framework.}
    \label{fig:toolArch}
\end{figure}

For Phase~2, we have created a bitstream converter and manipulation framework which uses the previously generated database. It is capable of converting a targeted bitstream back to its partial XDL netlist representation, e.g., routing, flip-flops, MUXs, and LUTs. Additionally, it is capable of modifying LUTs and PIPs directly in the bitstream, e.g., it can manipulate the Boolean LUT equations and set/unset PIPs in any arbitrary switch matrix. Figure~\ref{fig:toolArch} shows the architecture of our developed tool, which is written in \emph{C++}.
During development, we used a modular architecture allowing to support new FPGA devices and features, e.g., to support the conversion of BRAM or IO blocks.

We describe all modules as follows.
\begin{itemize}
 \item The database wrapper parses the previously created database and provides an interface to it. It stores all information regarding the mapping of bitstream bits to a hardware configuration.
 \item Similarly, the bitstream wrapper parses the targeted - to be converted - bitstream. The targeted bitstream is XORed with a generated reference bitstream that encodes an empty hardware configuration. This eliminates unwanted default configuration bits.
 \item The hardware configuration information, e.g., a list of PIP objects, is stored into the hardware configuration wrapper, which can be accessed by the XDL writer.
 \item The XDL writer is responsible for generating valid XDL code from the available information of the hardware configuration wrapper.
\end{itemize}

In the following, we discuss the bitstream reverser class in greater detail, as it is mainly responsible for correctly extracting the hardware configuration from a targeted bitstream.
Algorithm~\ref{aspdac:BSConversion} describes the entire extraction workflow.

\SetAlFnt{\small}
\begin{algorithm}[h]
	\caption{Bitstream conversion algorithm needed for extracting PIP configurations from a targeted bitstream.}
	\label{aspdac:BSConversion}
	\KwData{bitstreamWrapper* \textit{bs}, database* \textit{db}, configurationWrapper* \textit{c}}
	\Begin {
		\tcc{Iterate over every bit of the bitstream}
		\For{bitPos = bs$-$>getSyncWordPos() to bs$-$>bitstreamLength()}{
			\tcc{Check if the bit is set and marked}
			\If{bs->getBit(bitPos) == 1 \textbf{and} !marked}{
				\tcc{Search for a suitable object in the db}
				DataObject* \textit{obj} $\leftarrow$ \textit{db}$-$>GetElementFromDatabase(\textit{bitPos})\;
				\If{obj != nullptr}{
					\tcc{If an object is found, process it further depending on its type}
					\textit{type} $\leftarrow$ \textit{obj}$-$>getType()\;
					\uIf{type == PIP}{
						\tcc{Reverse the PIP}
						mostBitsToggled $\leftarrow$ 0\;
						usedPIP $\leftarrow$ nullptr\;
						\For{PIP p : db$-$>getLinkedPIPs(bitPos) }{
							\If{bs->bitsToggled(p$-$>getToggledBits())} {
								\If{mostBitsToggled < p$-$>getNumToggledBits()}{
									mostBitsToggled $\leftarrow$ p$-$>getNumToggledBits()\;
									usedPIP $\leftarrow$ p\;
								}
							}

						}

						\If{usedPIP != nullptr} {
							\tcc{set usedPIP in the config}
							c$-$>addPIP(p)\;
						}

					}
					\uElseIf{type == lutBit}{
						\tcc{Reverse the LUT Bit}
					}
				}
			}
		}
	}
\end{algorithm}

As can be seen, the algorithm iterates over every single set bit from the targeted bitstream for which the bitstream wrapper removed the disturbing default bits in a previous processing~step.

Depending on the bit's position, our database assigns it to the correct bits' object type, e.g., PIPs, MUXs, or LUTs as well as to the correct location on the two-dimensional FPGA grid. Then it stores the returned information in the hardware configuration wrapper.

Since the reconstruction of LUTs is for example already explained in~\cite{4101017,TCAD,rannaud2008bitstream}, we only describe the reconstruction of PIPs in further detail, as its conversion to a correct PIPs' object list requires appropriate processing.

\begin{figure}[htb]
	\centering
	\begin{subfigure}[b]{\columnwidth}
		\includegraphics[width=\columnwidth]{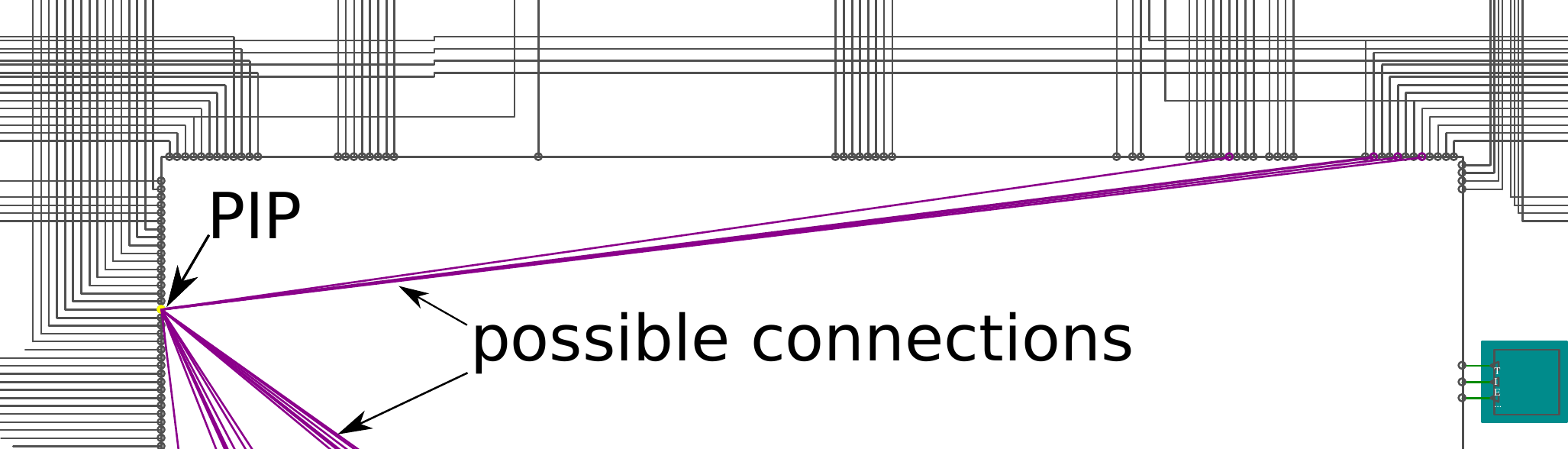}
		\caption{With all possible connections for one sink}
	\end{subfigure}
	\begin{subfigure}[b]{\columnwidth}
		\includegraphics[width=\columnwidth]{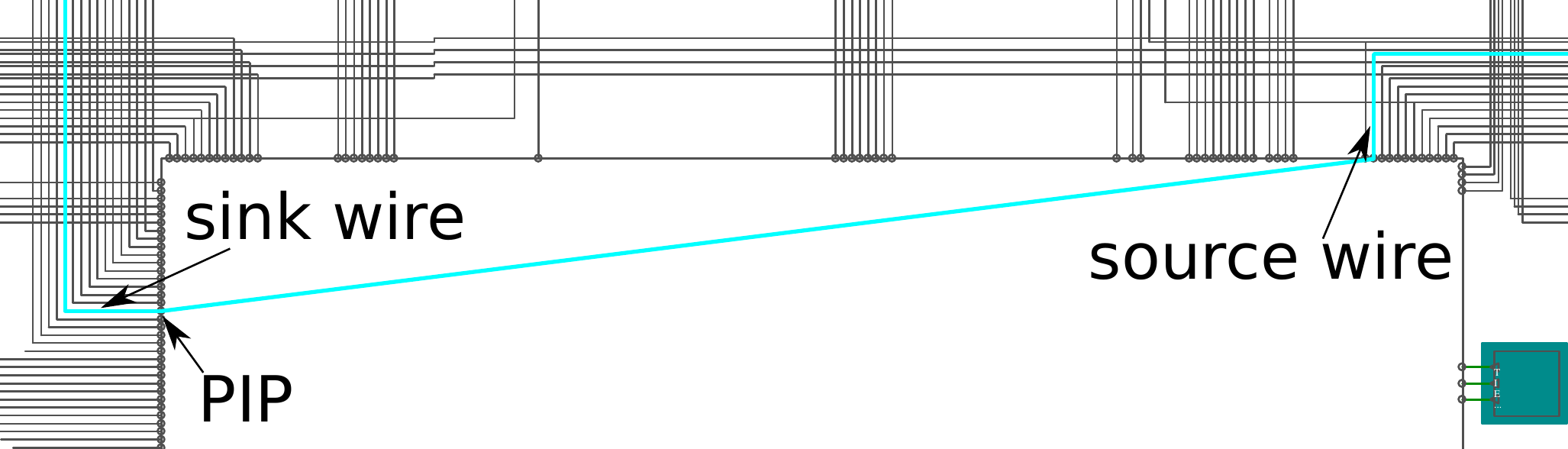}
		\caption{With one configured PIP}
	\end{subfigure}
	\caption{Upper part of a switch matrix}
	\label{Fig:aspdac:sink_conns}
\end{figure}

A switch matrix of an FPGA contains multiple sinks and sources. From now on, we only consider one sink, e.g., a PIP. Depending on the type, $N$~configurable sources are wired with the PIP, cf.~Fig.~\ref{Fig:aspdac:sink_conns}, but only one valid PIP connection can be routed through the sink at the same time, cf.~Figure~\ref{Fig:aspdac:sink_conns}.

Roughly speaking, a PIP acts like a multiplexer. The configuration of which source is routed through the PIP is stored in the bitstream. Such an N-to-1-multiplexer requires at least $\ceil{log_2(N)}$ control bits, but we observed that the bitstream stores more than $\ceil{log_2(N)}$ bits (without further explanation).
The fact that a single PIP share multiple configuration bits and these bits are not stored continuously complicates the bitstream conversion task. To address this, we have developed Algorithm~\ref{aspdac:BSConversion}.

\begin{figure}[htb]
    \centering
    \includegraphics[width=1\columnwidth]{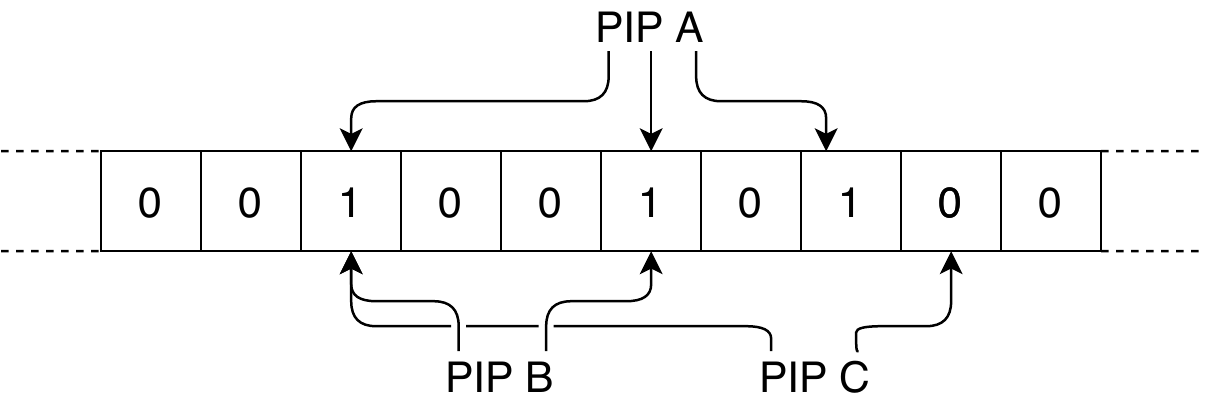}
    \caption{Possible distribution of PIP's configuration bits within a bitstream. If all arrows of a PIP point to only set bits and if it has the most bits set, the corresponding PIP is configured. If any outgoing arrow of a PIP points to at least one cleared bit, then the PIP is obviously not configured.}
    \label{fig:toggledPIPs}
\end{figure}
As indicated before, our tool iterates over each bit in the targeted pre-processed bitstream. If the algorithm encounters a PIP configuration bit, it yields a set of possible PIPs and iterates over each PIP candidate. It then successfully figures out, which PIP is the correct candiate, cf. the caption of Figure~\ref{fig:toggledPIPs} or Algorithm~\ref{aspdac:BSConversion}.

In the shown example of Figure~\ref{fig:toggledPIPs}, our tool processes the most-left set PIP configuration bit. In this case, the database will return PIP~A, PIP~B, and PIP~C as potential configured PIPs. Since one bit for PIP~C is cleared, our algorithm discards this candidate. As a valid PIP requires to have set all configuration bits. Even though this is the case for PIP~B, it will be discarded as well, since all configuration bits of PIP~A are set and the Hamming weight of PIP~A is the largest one. Therefore, in this toy example PIP~A is the correct PIP candidate.

Additionally, once we have reconstructed all configured PIP bits within a bitstream, we convert back all set NETs. For this purpose, we have re-implemented the described ICG algorithm from Ding~\textit{et al.}~\cite{DingWZZ13}.



\subsection{Bitstream Manipulator}
Note that we started to implement a bitstream manipulator to avoid the need for a complete bitstream conversion. So far, our framework is capable of setting or unsetting PIPs and changing Boolean functions in LUTs.
This allows us to carry out targeted bitstream manipulations for which we do not possess the fully reverse engineered XDL netlist. Therefore, we do not rely on the vendor's tools, e.g., xdl and bitgen. They usually require a complete XDL netlist to ensure the intended functionality of a hardware configuration. In the future, we plan to enhance our bitstream manipulation capabilities further.

Having introduced our capabilities, we now describe how a Trojan may be inserted into a targeted selftest-protected AES IP core.

\section{Case Study: Trojan Insertion}
\label{hal:section:case_study}
Only little is known about how a Trojan designer may exactly proceed to tamper a third-party hardware configuration. This case study demonstrates a possible Trojan insertion methodology into an \ac{AES} IP core at hardware configuration level. Note that in practice such Trojan requires to either develop or use a tool that can

\begin{itemize}
    \item convert the bitstream file format to a gate-level netlist, cf.~Section~\ref{aspdac:sec:intro}~and~\ref{aspdac:sec:bs_rev}
    \item manipulate the underlying hardware primitives without violating any timing constraints.
    \item perform correct bitstream file patching
\end{itemize}

In the field, a bitstream manipulation is possible during various life-cycle phases of a device, i.e., it can be intercepted during shipment, during selling, or even during operation from the customer himself. Even though we did not test the following manipulations directly on the bitstream, we provide a new idea of how an attacker can proceed to manipulate a placed and routed hardware configuration where only partial information is available, i.e.., we assume that a third-party bitstream was partially converted. This is a realistic assumption\footnote{It is likely that state actors are able to obtain all data from a bitstream, but we nevertheless try to explore the actual practical capabilities of weaker attackers who only possess partial information} for most deployed embedded devices, where the bitstream is usually stored in plaintext on the same PCB along with the targeted FPGA. We used the \textit{HAL} framework~\cite{HAL} to identify and manipulate the relevant netlist components. From our result, we are confident that a hardware configuration must not be fully known to also accomplish more complicated Trojan insertions at a very late stage.

\subsection{Target and Notation}
Our targeted AES-128 IP core provides an interface to set a key for data decryption and encryption. To interact with the circuit we integrated an \ac{UART}/RS-232 interface. Further, we use the following notations:
$p$ - Plaintext (16 bytes), $k$ - Key (16 bytes), $c = \mbox{AES}_k(p)$ - Ciphertext (16 bytes), $(p_{\text{ref}}, c_{\text{ref}})$ - Plaintext/ciphertext pair for the self-test, $k_{\text{st}}$ - Key for the self-test, $k_u$ - Key for user data.

\subsection{\bf System Model}
We assume that an \ac{FPGA} and an external entity, e.g., a dedicated microprocessor, are integrated on the same embedded system along with an \ac{FPGA}-based AES accelerator in a bitstream. The workflow is as follows:
\begin{enumerate}
    \item After the \ac{FPGA} is configured with the AES IP core, the external entity conducts a self-test with the FPGA by setting the self-test key $k_{\text{st}}$ and the reference plaintext $p_{\text{ref}}$.
    \item The external entity analyzes the FPGA's computed ciphertext $c= AES_{k_{\text{st}}}(p_{\text{ref}})$ and verifies the integrity by comparing if $c$ is equal to the reference ciphertext~$c_{\text{ref}}$.
    \item In case the self-test is successful, an AES key $k_u$ is derived after a user legitimates to unblock his encrypted device. A key derivation is computed and the user key is automatically passed to the FPGA.
\end{enumerate}
Note that such an embedded system is not only a theoretical assumption, since it is similar to the FIPS-140-2 level 2 certified USB flash drive from Kingston, cf. the work of Swierczynski~\textit{et~al.}~\cite{USBK}.

\subsection{Adversary's Goal and Trojan Idea}
The high-level goal of the adversary is to permanently burn the key $k_{\text{st}}$ into the FPGA hardware configuration, so that the user-data is always encrypted/decrypted with $k_{st}$ regardless of which key $k_u$ is set by the user. This way, the adversary can, later on, decrypt all seemingly securely encrypted user data. To accomplish this goal, the adversary targets to attach a payload circuit to the existing set-key circuit to permanently write the known key bits of $k_{st}$ into the \acp{FF} that process~$k$. Thus, the external self-test falsely confirms the integrity of the \ac{FPGA} \ac{AES} \ac{IP} core. To sum up, the described Trojan can trick an external self-test in cases where the self-test key is known like in~\cite{USBK}.

\subsection{Detection of Set-Key Circuitry}
\begin{figure}[!htb]
    \centering
    \includegraphics[width=1\columnwidth]{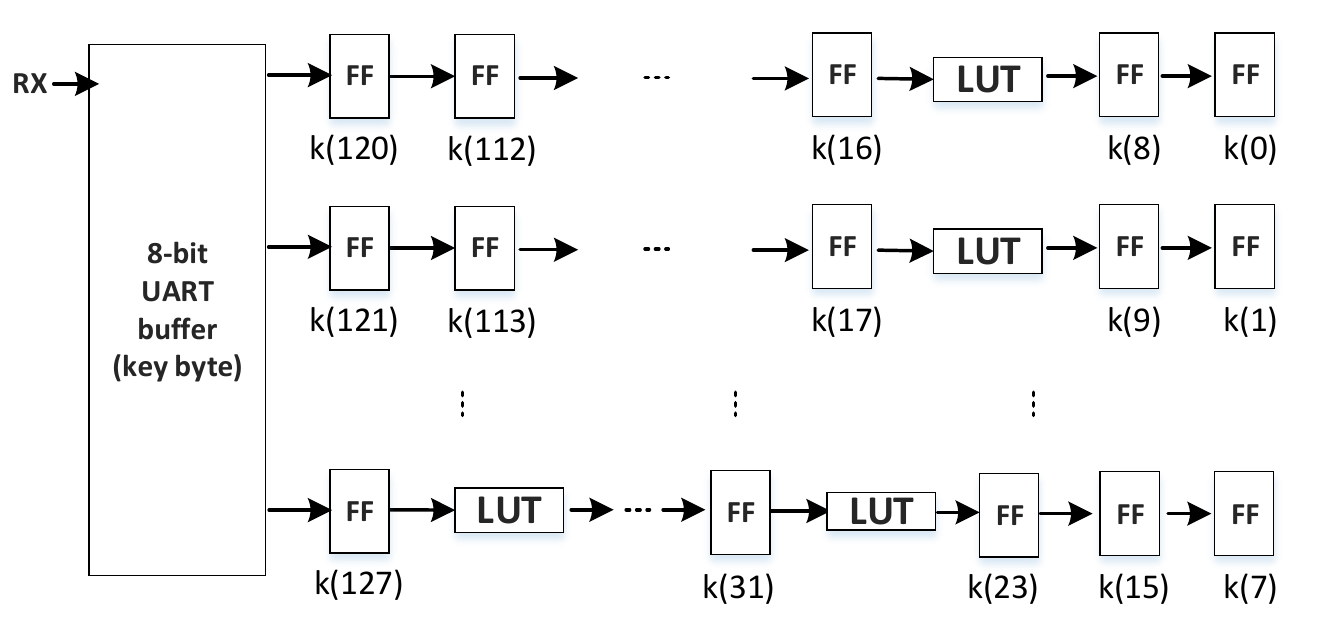}
    \caption{8 observed parallel 16-bit shift-registers sampling one key byte from a repeatedly updated 8-bit buffer.}
    \label{figure:shift-register-normal-byte}
\end{figure}

For our targeted \ac{AES} core, we noticed that eight 16-bit shift-registers were integrated into the hardware configuration. Its task is to temporarily store the key received sequentially via \ac{UART} and is later on forwarded to the \ac{AES} IP core, cf. Figure~\ref{figure:shift-register-normal-byte}. For an attacker, the key registers are an appealing target, as its alteration can, for example, enable key extraction or Trojan insertion. Since our goal is to burn a fixed key into the design, we first extract the relation between a \ac{FF} and the \ac{AES} key bit and second override the corresponding parts of the hardware configuration that delivers the (usually correct) key bit values. For that purpose, we developed a plugin to algorithmically detect deployed shift-registers and its corresponding length, i.e., how many \acp{FF} are cascaded, within a netlist.

In order to reverse-engineer the relation between each \ac{FF} and key bit, we manipulate/clear the first pass-through LUT (in this case it was always present, not shown in Figure~\ref{figure:shift-register-normal-byte}) of a found shift-register in the hardware configuration. By means of simulation\footnote{Note that one may also manipulate the corresponding change in the bitstream and observe the resulting ciphertexts through querying the AES module with several known plaintexts. This way one can also conclude the relation between flip-flops and key bits without conducting any simulation. For the sake of simplicity, we used simulation.}, we let the hardware configuration compute a ciphertext for a known plaintext and an AES key for which all 128 key bits are set. By comparing the resulting FPGA ciphertext output with various pre-computed known plaintext/ciphertexts pairs for suitable keys\footnote{For example, we computed look-up tables with plaintext/ciphertext pairs for a key, where the last bit of each of 16 key bytes was cleared, while all other key bits were set.}, we can conclude the relation of each shift-register and the key bit position of a key byte it processes. Note that we similarly determined whether the first or last flip-flop of a shift-register store the key bit of the first or last key byte. Having figured out which hardware primitive is responsible for processing one segment of the AES key, we can now describe the Trojan payload.

\subsection{Manipulating a Set-Key Circuitry}
The key idea is to detach any signal from the data input pin \textsc{d} of each key bit \ac{FF} and to attach the output signal from the so-called \textit{payload \acp{LUT}},~cf.~Figure~\ref{figure:shift-register-normal-byte}, so that all key-based shift-registers are independent from the key buffer.

For the communication with our device an RS232 interface is used, which receives the key byte by byte. Eight shift-registers are used to store the bits of the current key byte during the loading phase. Each shift-register itself can hold up to 16 values to store all 8-bit chunks of a 128-bit key ($key\_byte$ in Figure~\ref{figure:shift-register-normal-byte}).
In total, we added 128 payload \acp{LUT} and routed them to their target \acp{FF}, using \textit{HAL}. Of course alternative approaches may be used like for example attaching GND or VCC signals to the corresponding flip-flops inputs, but so far we did not reverse-engineer the corresponding bitstream encoding for those hardware elements. Each payload \ac{LUT} is programmed individually to output either $0$ or $1$ (depending on the key bit $k_{\text{st}}$). Overall, the functionality of the $AES_k(\cdot)$ is substituted by the malicious one, i.e., by $AES_{k=k_{\text{st}}}(\cdot)$.
\begin{figure}[!htb]
    \centering
    \includegraphics[width=1\columnwidth]{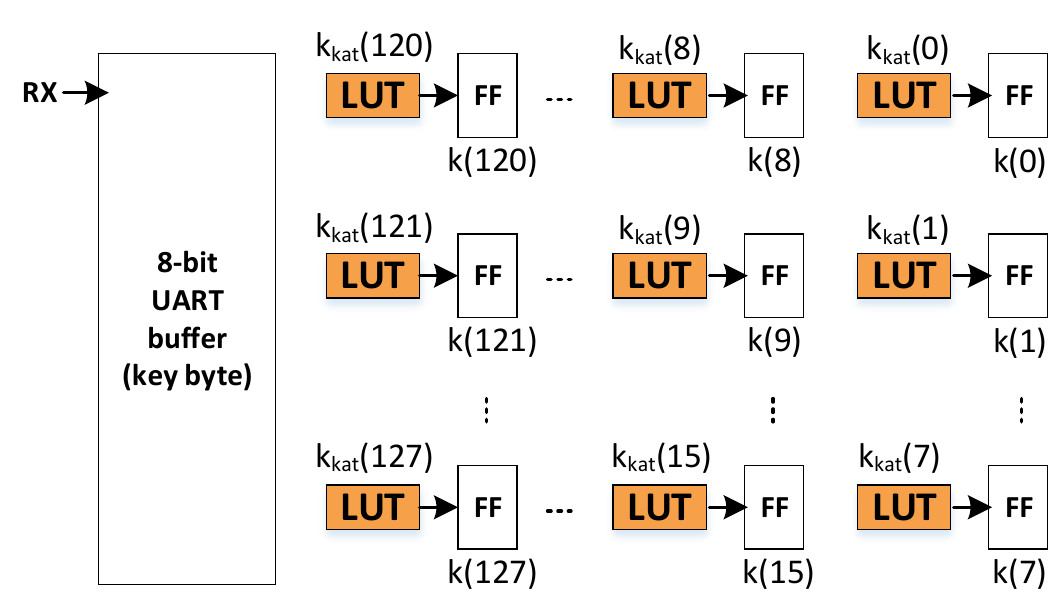}
    \caption{Key-based Shift-register after IP manipulation.}
    \label{hal:fig:shift-register-payloaded-byte}
\end{figure}

We verified the correctness of our Trojan by setting a random key $k \neq k_{st}$ and testing that the manipulated AES core nevertheless computes $AES_{k_{st}}(\cdot)$ when encrypting one plaintext.

\subsection{Stealthiness}
Considering this Trojan, the self-test is rendered useless if only one key $k_{\text{st}}$ is tested by the external entity. Furthermore, all user data is encrypted with the known key $k_{\text{st}}$ instead of the derived key $k_{u}$ making the decryption entirely possible.
Note that in case of an embedded device such as the \ac{USB} flash drive~\cite{USBK}, the user usually has no direct access to $k_u$ unless he attempts to derive $k_u$ by himself, which requires exact knowledge of the key derivation function. Certainly, once the user has access to $k_u$ and obtains one plaintext/ciphertext pair, this Trojan can be easily detected.

\section{Conclusion and Future Work} 

In this paper, we demonstrated further attackers' capabilities based on the example of a malicious manipulation of a self-test protected AES core. We improved the bitstream reverse engineering methods by simplifying known routing extraction mechanisms. Thus we conclude that the efforts for a successful bitstream conversion are even lower than commonly assumed. 
Note that bitstream reversing is also an essential step for legitimate purposes such as Trojan detection, \ac{IP} theft exposure, bitstream verification, or advanced bitstream tooling. 
Further, we demonstrated the implication of partial bitstream exposure. Combined with our Trojan case study, we emphasized that key-based shift-registers can be exploited. The identification of such points of interest is usually one of the various practical hurdles during reverse engineering. 
For future projects, one may also analyze how to algorithmically identify key registers with more complicated hardware structures and whether partial bitstream reverse engineering is sufficient to carry out similar attacks. Consequently, future research should explore defense mechanisms for such vulnerable hardware structures, e.g., hardware obfuscation methods. 
\section{Acknowledgement}
Part of this work was supported by the European Research Council (ERC) under the European Union's Horizon 2020 Research and Innovation programme (ERC Advanced Grant No. 695022 (EPoCH)) and by NSF grant CNS-1563829.

\bibliographystyle{ACM-Reference-Format}
\bibliography{bibliograpghie}

\end{document}